\documentclass[a4paper,11pt]{article}
\pdfoutput=1
\usepackage{jinstpub} 
\usepackage{xcolor,soul,framed} 
\usepackage{graphicx}
\usepackage{subcaption}
\usepackage{wrapfig}
\usepackage{mdwmath}
\usepackage{mdwtab}
\usepackage{eqparbox}
\usepackage{url}
\usepackage{dblfloatfix}    
\usepackage{hyperref}
\usepackage[export]{adjustbox}

\begin{document}
    \title{Unsupervised Learning Methods in X-ray Spectral Imaging Material Segmentation}
    
\author[a,1]{Jericho~O'Connell,\note{Corresponding author.}}
\author[a]{Kevin J. Murphy}
\author[a]{Spencer M. Robinson}
\author[b]{Kris~Iniewski}
\author[a]{Magdalena~Bazalova-Carter}

\emailAdd{jerichoo@uvic.ca}

\affiliation[a]{University of Victoria,\\3800 Finnerty Rd, Victoria, BC, Canada}
\affiliation[b]{Redlen Technologies,\\123 - 1763 Sean Heights
Saanichton, BC, Canada}


\abstract{
In this work, we have investigated a number of unsupervised learning methods for material segmentation in projection x-ray imaging with a spectral detector. A phantom containing two hard materials (glass, steel) and three soft materials (PVC, polypropylene, and PFTE) all embedded in PMMA was imaged with a 5 energy bin spectal detector. The projection images were utilized to test nine unsupervised learning algorithms for automated material segmentation. Each algorithm was investigated using single energy (SE), dual energy (DE) and multi energy (ME) images. Clustering results were scored based on homogeneity and completeness of the clusters, which were combined into the Rosenberg and Hirshberg's V-measure. Principle component analysis (PCA), independent component analysis (ICA), and non-negative matrix factorization (NMF) were tested as dimensional reduction methods. ME, DE and SE material segmentation was performed using five, two, and single energy images, respectively. ME had the highest V-measure on the soft materials using PCA and a novel interpolating bayesian gaussian mixture model (BGMM) clustering with a V-measure of 0.71. This was by 3.5\% better than DE and 20.3\% better than SE. Conversely, SE imaging was most capable of hard tissue segmentation using the standard BGMM, with a V-measures of 0.84. This was 6.3\% better than DE and 5.0\% better than ME. This work demonstrated that ME x-ray imaging might be superior in segmenting soft tissues compared to conventional SE x-ray imaging.

}

\maketitle
%


\section{Introduction}

Over the past century, radiography has remained the primary tool in diagnostic imaging. Due to its frequent use, radiography gives a large radiation dose to the general population. Additionally, as a front line diagnostic tool, inaccuracies in radiography are responsible for false-positive findings. In turn, these findings subject healthy patients to more radiation and increase health care costs. For example, mammography's positive predictive power is estimated to be as low as 20\% \cite{Skaane2013ProspectiveArbitration., Dickersin2010TheCancer, Kopans1992TheMammography., Mushlin1998EstimatingMeta-analysis, Chiarelli2013DigitalProgram} due to a high rate of false positives. Longer term, false-positives are reported to be as high as 60\% for women undergoing screening mammography over 10 years \cite{Kerlikowske2013OutcomesTherapy, Hubbard2011CumulativeMammography}. Thus, we look at new x-ray imaging modalities such as spectral or multi-energy (ME) and dual energy (DE) radiography as means to improve the false-positive rates or provide similar image quality with a lower imaging dose.


Recently, Monte Carlo (MC) simulations have shown added benefits in tissue differentiation tasks using ME CT \cite{Lalonde2016ACT} thus indicating the benefit of multi-energy imaging. Further, in planar imaging, which is the subject of this work, the use of more than two energies has been shown to improve material differentiation \cite{OConnell2019OptimalDetector}: Planar images typically contain an unknown depth of material, which must also be determined to identify the material. Thus ME image is hypothesized to aid in planar imaging segmentation tasks. Since ME and DE imaging are concerned with differentiation of soft materials based on differences in the x-ray beam attenuation rather than the mass density, new methods are applied to analyze materials. Much of the work in DE CT focused on the calculation of elemental compositions for MC dose calculations and thus converts the DE images into the electron density and effective atomic number to characterize the material \cite{Bazalova2008Dual-energyCalculations,Landry2013DerivingCoefficients,Saito2017ABody}. These calculations assume the uniformity of each CT voxel, reasonable for CT, for planar imaging the region of interest will be non-uniform. This leaves room to explore new methods for segmentation with this modality.

In this work image segmentation using a Cadmium Zinc Telluride (CZT) ME detector with five energy bins is explored. Recently there has been increased interest in multi-energy x-ray imaging modalities, however other imaging modalities, like hyperspectral imaging, have a long history of high dimensional image analysis \cite{Khan2018ModernReview}. Hyperspectral imaging is performed at photon energies near the visual spectrum and sees application in very different use cases than DE x-ray imaging \cite{Lu2014MedicalReview.,Khan2018ModernReview}. However the data is in the format of a multi-dimensional image, with each dimension acquired at a seperate energy, this has parallels to ME imaging. In hyperspectral imagining clustering methods are successfully used to automatically segment images \cite{Murphy2018UnsupervisedDiffusion,Gillis2012HyperspectralGraphs,Noe2001PartialClustering}. Likewise dimensional reduction is standard in the pre-processing of the images \cite{Mahesh2015HyperspectralMaterials}. With these methods as motivation we apply clustering methods to ME imaging as well as single energy (SE) and DE imaging.

In SE CT imaging clustering methods have applied for segmentation. Yoa et al. applied a gaussian mixture model (GMM) to segment spinal CT volumes for automated segmentation in a surgical setting \cite{Yao2004ColonicModels}. Likewise, Li et al. apply a fuzzy c-means segmentation in order to segment colon polyps \cite{Li2008ImprovedRadiography}. However to the best of the authors' knowledge a comprehensive comparison of clustering methods on an x-ray imaging task has not been performed. This work aims to be the first general comparison of clustering methodologies on a planar imaging task. This work also aims to compare performance of SE, dual, and ME clustering methods using data acquired with a CZT ME detector with five energy bins in a phantom study. Further, a novel interpolating clustering algorithm is proposed for optimal soft tissue segmentation.


\section{Materials and Methods}
\label{sec:methods}

\subsection{Data Aquistion}

\begin{figure}[htbp]

\includegraphics[width=\textwidth]{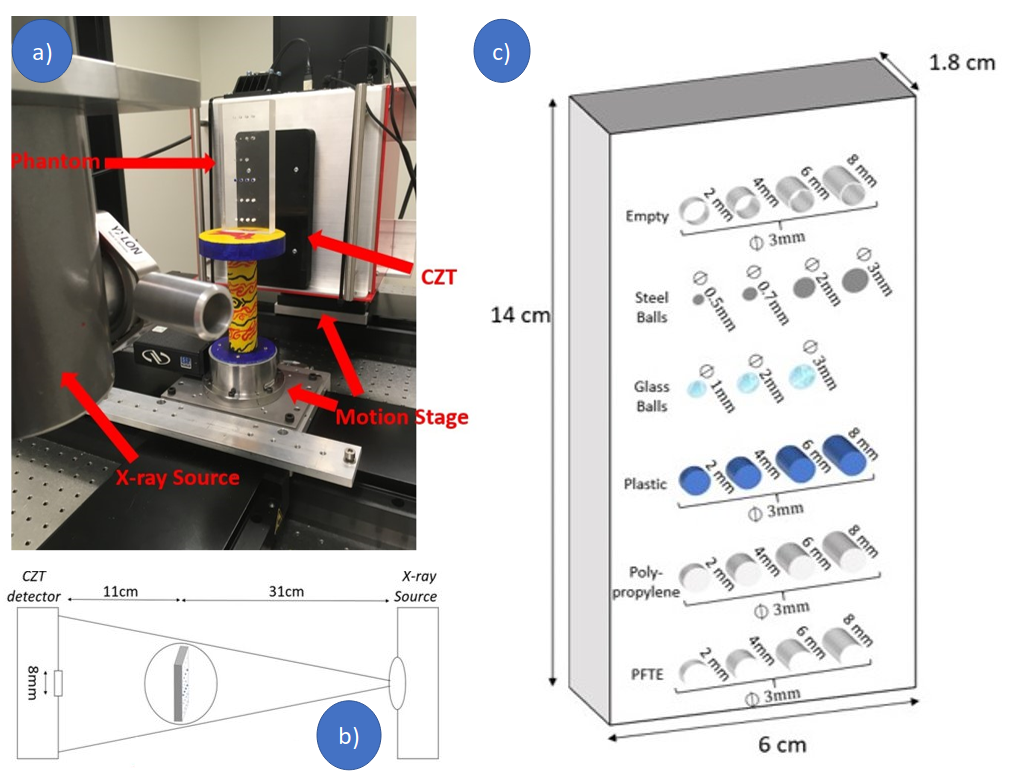}

  \caption{a) A photograph and b) a schematic of the imaging setup. c) A schematic drawing of the PMMA phantom used for material segmentation tests.}
  \label{figure:setup}
\end{figure}

Planar x-ray images of a rectangular PMMA phantom with five materials (steel, glass, PVC, polypropylene, and PTFE) were acquired with a CZT detector (Redlen Technologies, Victoria, BC, Canada) on a table-top imaging system. PVC, polypropylene and PTFE were considered to be the soft materials while glass and steel were considered the hard materials in our material segmentation analysis. These materials were used as a benchmark as they are cheaply available and span the range of densities found in the human body (0.95 - 1.92) \cite{White1989Report44}. Steel is also used as it is commonly found in medical imaging. Densities and effective atomic numbers of these materials can be seen in Table \ref{table:densities}. PTFE was considered a soft material, although denser than bone, it has an effective atomic number similar to soft materials and provided contrast similar to the soft materials in PMMA. The effective atomic numbers were calculated using the method of Murty et al. \cite{MURTY1965EffectiveMaterials}.

\begin{table}[b]
\caption{Material Densities}
\begin{tabular}{lllllll}

                                                       & PMMA & PP    & PVC   & PTFE & Glass & Steel \\ \hline
Density {[}g/cm$^3${]}                  & 1.18 & 0.94  & 1.11  & 2.1  & 2.8   & 8.0   \\
Effective Atomic Number & 6.56    & 5.35 & 14.3 & 8.5 & 18.1 \cite{Sharma2012EffectiveGlasses}  & 26 
\end{tabular}
\label{table:densities}

\end{table}

Data was acquired using a CZT array capable of sorting x-rays into six energy bins with a 8$\times$12 mm imaging array. The 330 $\mu$m pitch high-flux CZT detector is 2mm thick \cite{Thomas2017CharacterisationCdZnTe,Iniewski2014CZTImaging}. Travel Heat Method (THM) was adopted when growing the CZT crystals were placed in a sensor that was connected to a photon counting ASIC. This detector communicated with an external PC though Gigabit I/Os to a PC. The energies of photons incident with the detector are sorted into five energy bins by the ASIC. In the case of this experiment the energy bins were set to 16-33 keV, 33-41 keV, 41-50 keV, 50-90 keV, and 90-120 keV.

The detector and X-ray source (XRS-160, Comet Technologies, Bern, Switzerland) were both mounted on vertical and horizontal linear motion stages (Newport technologies, Irvine, CA). These stages were oriented perpendicular to each other to allow for easy navigation while imaging the phantom, which was placed on a stage in between the x-ray source and the CZT detector. The x-ray source used was a module XRS-160 from Comet Technologies. A photograph of the experimental setup can be seen in Figure \ref{figure:setup} a). 

The PMMA phantom block in Figure \ref{figure:setup} c) was placed on the stage and the 3 smallest inserts of each material were imaged. Air scans were also acquired for image normalization as $-log\frac{I}{I_o}$ where $I$ is the image intensity and $I_o$ is the air scan. During all scans the X-ray tube was using a cone beam operated at 1 mA tube current, 120 kV tube voltage with a 1-mm focal spot. A DE image was created for each ME image by summing over the first two (16-41keV) and last three 
(41-120 keV) bins of the image while a SE image was created by summing over all of the bins.

\subsection{Image Segmentation}
\subsubsection{Overview}

\begin{figure}[htbp]

\includegraphics[width=\textwidth]{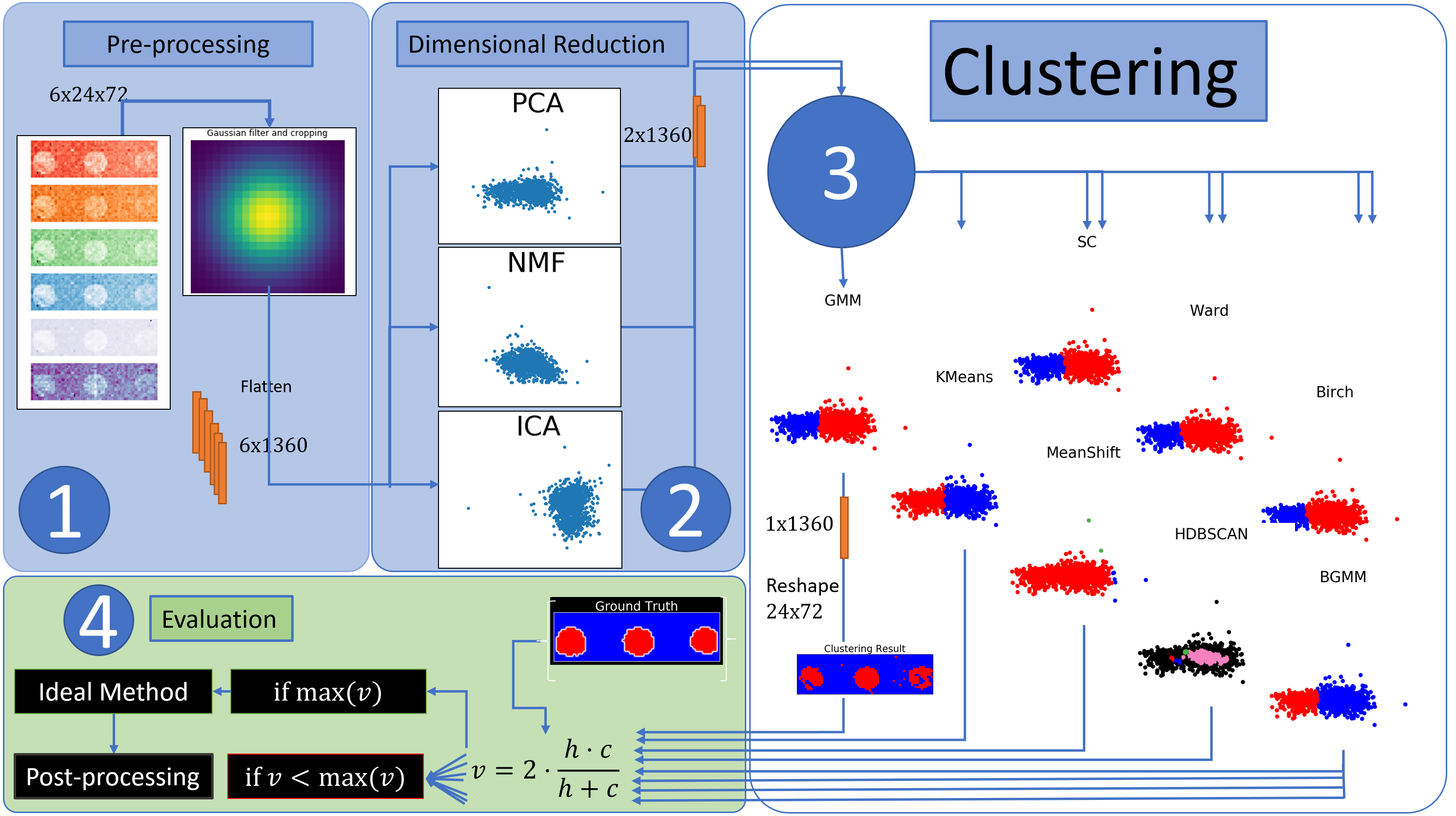}

\caption{A general overview of the workflow outlines the four general steps of the segmentation method.}
\label{overview}
\end{figure}

\noindent A complete overview of the analysis and segementation workflow is depicted Figure \ref{overview}. Image segmentation was undertaken as a multistage process. Here we applied techniques used in hyperspectral imaging; Mahesh et al. \cite{Mahesh2015HyperspectralMaterials} state that the main steps in hyperspectral images segmentation include:
\begin{itemize}
\setlength\itemsep{0em}
\item Pre-processing of data 
\item Dimensional reduction
\item Enhancement of spectral responses
\item Component detection or classification 
\end{itemize}
These steps were also applied in our work and they are summarized in Figure \ref{overview}. The enhancement of spectral response is not a step at this time, as the modelling of the spectral response of the CZT detector is beyond the scope of this study.

\subsubsection{Pre-processing of Data}

\begin{figure}[htbp]

\includegraphics[width=\textwidth]{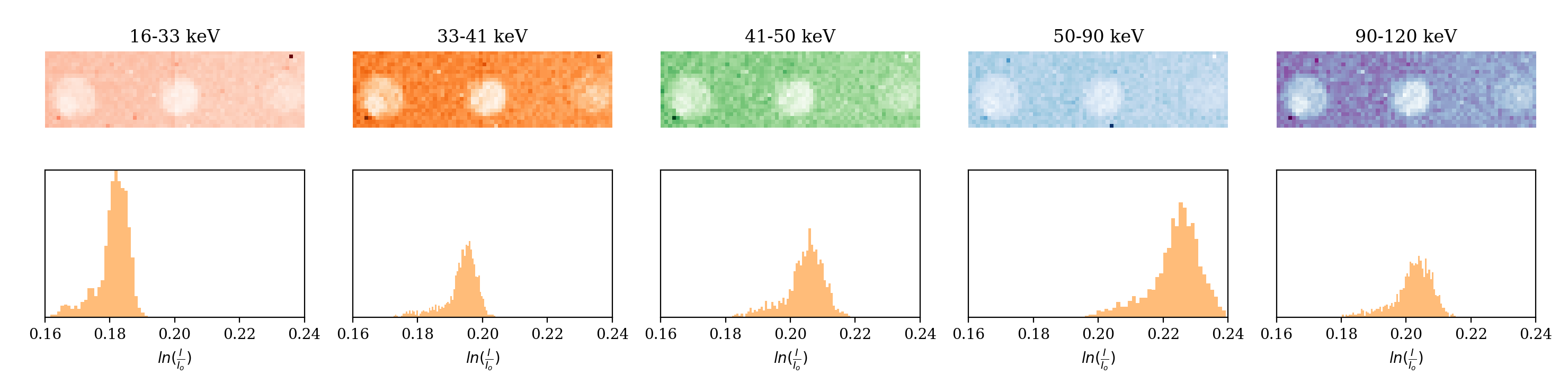}

\caption{The six energy bins for the image of PMMA embedded with poly-propylene are shown with different colormaps corresponding to their energy, red being lower energy while violet is higher energy. A histogram of the log intensity corresponding to each image is shown below the respective image.}
\label{demonstrating_bins}
\end{figure}

Data pre-processing was performed using MATLAB 2017b (The MathWorks, Natick, USA). The images were first cropped to remove the non-uniform edge pixels of the detector. Dead pixels interpolated to be the average of the surrounding eight pixels. The images were then smoothed using a two dimensional gaussian filter with a standard deviation of 0.115 $\mu$m to reduce the noise in the image. The five input images after the initial pre-processing are shown in Figure \ref{demonstrating_bins}.

\subsubsection{Dimensional Reduction}

\begin{wrapfigure}{R}{0.5\textwidth}
  
  \begin{center}
    \includegraphics[width=0.48\textwidth]{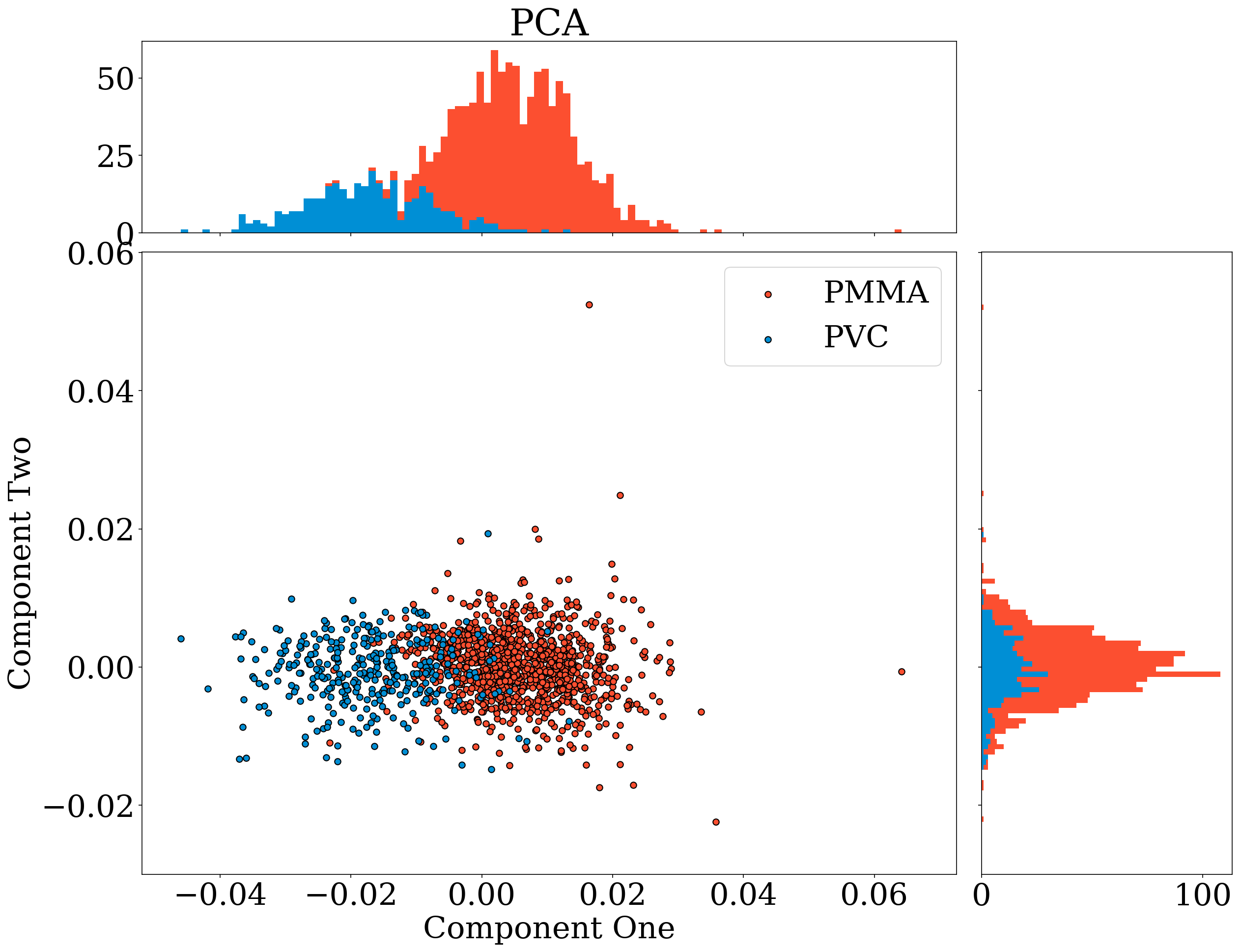}
  \end{center}
  
  \caption{The first two principle components of the PTFE image displayed as a scatter plot.}
  \label{PCA}
  
\end{wrapfigure}

Dimensional reduction methods were applied in this work to increase class separation and reduce noise in the data, Python (Python Software Foundation) using sci-kit learn \cite{Pedregosa2011Scikit-learn:Python}. The three dimensional reduction tehcniques considered suitable for this task were principle component analysis (PCA), independent component analysis (ICA), and non-negative matrix factorization (NMF).

\subsubsection{Principal Component Analysis}

PCA decomposes the covariance matrix of the feature space into eigenvalues and eigenvectors, eigenvectors are sorted in terms of the magnitude of their eigenvalue to find the directions of highest variance in the data. The data is then projected into a lower dimensional orthogonal space defined by the eigenvectors with the largest eigenvalues. This method results in a loss of information, however this loss of information is usually relatively small and ideally the discarded dimensions in the data amount to noise. PPCA example from this work applied on an image of PVC is shown in Figure \ref{PCA}. PCA is fast, linear and sees application in many domains. Jolliffe et al. \cite{Jolliffe2016PrincipalDevelopments.} present a more in depth overview of PCA and its recent applications.

  
  
  

\subsubsection{Independent Component Analysis}

Used for blind source seperation in time series analysis, independant component analysis (ICA) separates a mixed signal into its constituent signals. ICA is also used in hyperspectral imaging \cite{Villa2009OnAnalysis}. Using ICA we frame the segmentation of the two images as a decomposition problem in which the image is a weighted addition of two signals. Idealy these signals would be the background material (PMMA) and the material of interest (eg. PVC). A demonstration of using ICA on 
PVC embedded in PMMA can be seen in Figure \ref{ICA}. A more in depth explanation of ICA can be found in appendix. The algorithm used for ICA in this work was the FastICA \cite{Hyvarinen2000IndependentApplications} algorithm implemented in sci-kit learn.

\begin{figure}[htbp]
    \centering
    \begin{subfigure}[b]{0.48\textwidth}
        \includegraphics[width=\textwidth]{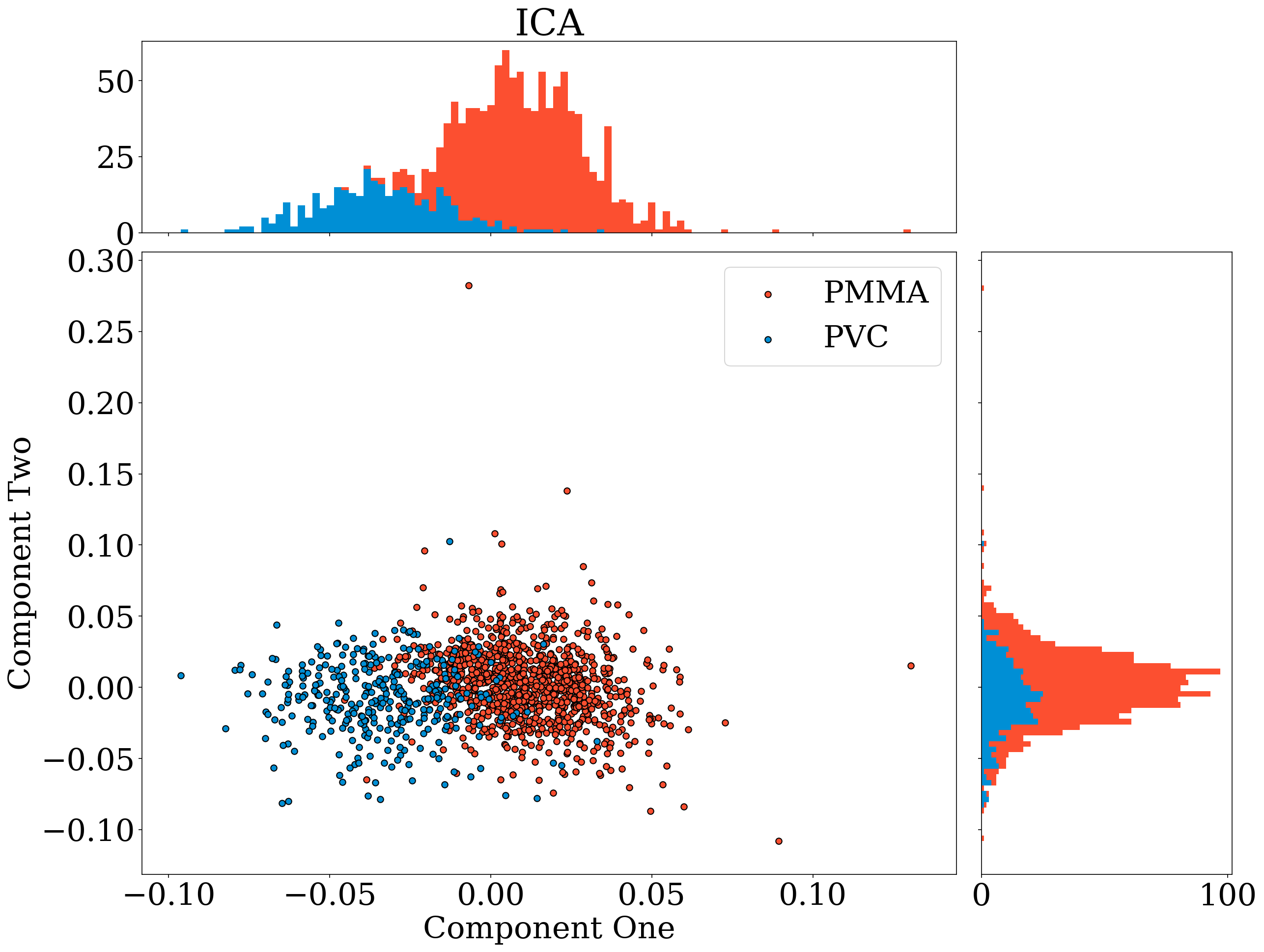}
    \caption{The first two independent components of the PTFE image displayed as a scatter plot.}
  
    \label{ICA}
    \end{subfigure}
    ~ 
    \begin{subfigure}[b]{0.48\textwidth}
        \includegraphics[width=\textwidth]{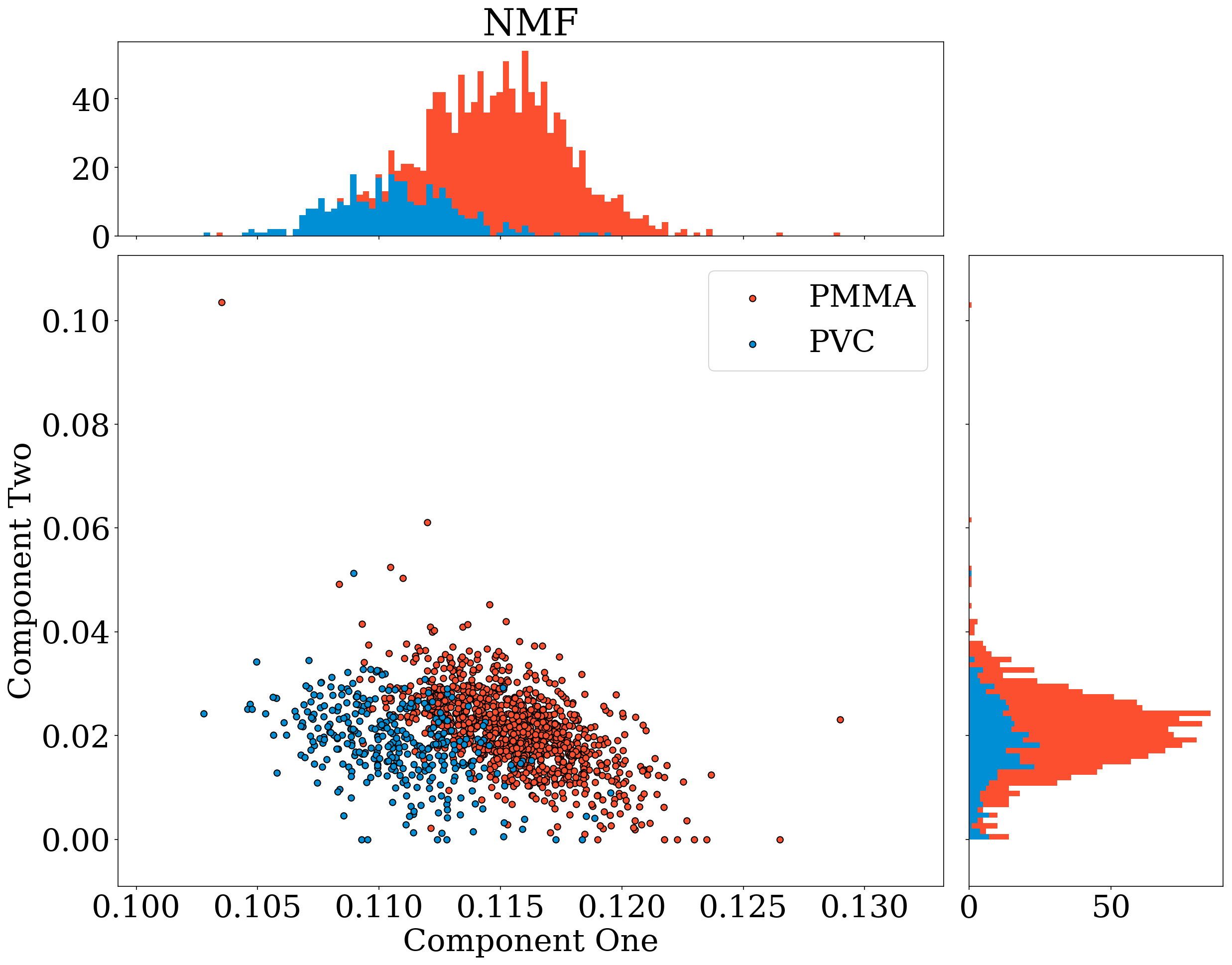}
    \caption{The first two non-negative factorization components of the PTFE image displayed as a scatter plot.}
  
    \label{NMF}
    \end{subfigure}
  \caption{ICA and NMF}
  
  \label{NMF_ICA}
\end{figure}

\subsubsection{Non-negative Matrix Factorization}

NMF assumes the Matrix $V$ is the product of the two matrices $W$ and $H$ such that : $V = WH$. 

When multiplying matrices the dimension of the factor matrices $W$ and $H$ may be much less that the product matrix. One can think of the two factor matrices as a feature matrix $W$ and a coefficient matrix $H$.

Thus, if one can find matrices $W$ and $H$ that have fewer dimensions than $V$, one can reconstruct the matrix $V$ in the lower dimensional space defined by $W$. To find the matrices $W$ and $H$ numerically we try to minimize the error defined by:

\begin{equation}
\min_{W,H} || V - WH ||, ( W \geq 0, H \geq 0 )
\end{equation}

This was done in using scikit-learn's NMF function, minimizing using the gradient descent algorithm. A demonstration of NMF on PTFE embedded in PMMA can be seen in Figure \ref{NMF}.

\subsection{Clustering Methods}

\subsubsection{K-means}

K-means clustering was implemented using the k-means++ algorithm \cite{ArthurK-means++:Seeding}. K-means is a general purpose, fast and scalable clustering algorithm. Different distance metrics were used, however none of the distance metrics tested improved performance over the default squared euclidean distance, thus squared euclidean distance was used throughout the manuscript. Like many of the methods examined K-means requires the parameter $K$ which describes the number of clusters in the data. This parameter can be difficult to estimate in some cases. Methods for defining the parameter $K$ will be described in the following metrics section.

\begin{figure}[t!]
    \centering
    \begin{subfigure}[b]{0.48\textwidth}
        \includegraphics[width=\textwidth]{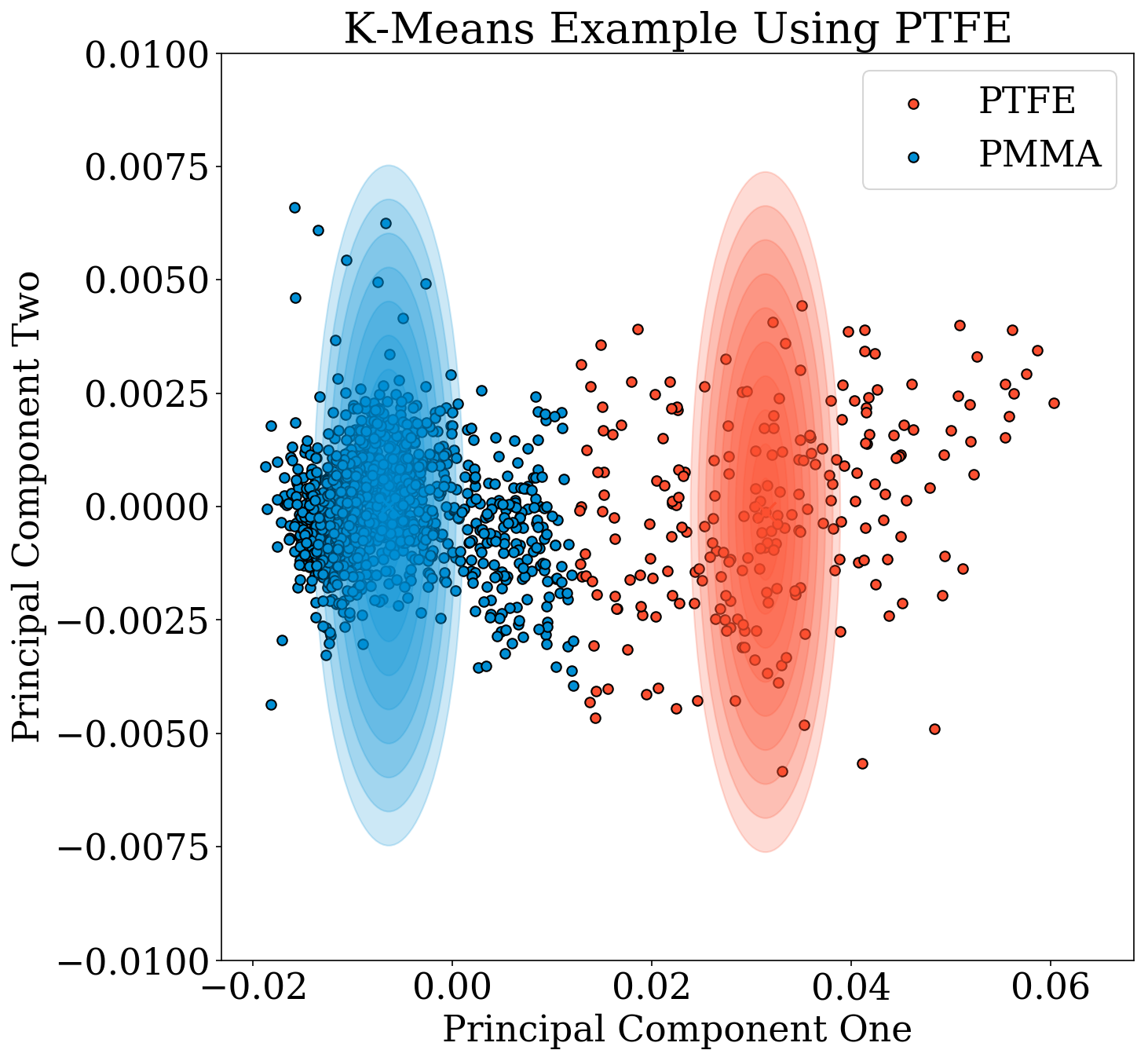}
    \end{subfigure}
    ~ 
    \begin{subfigure}[b]{0.48\textwidth}
        \includegraphics[width=\textwidth]{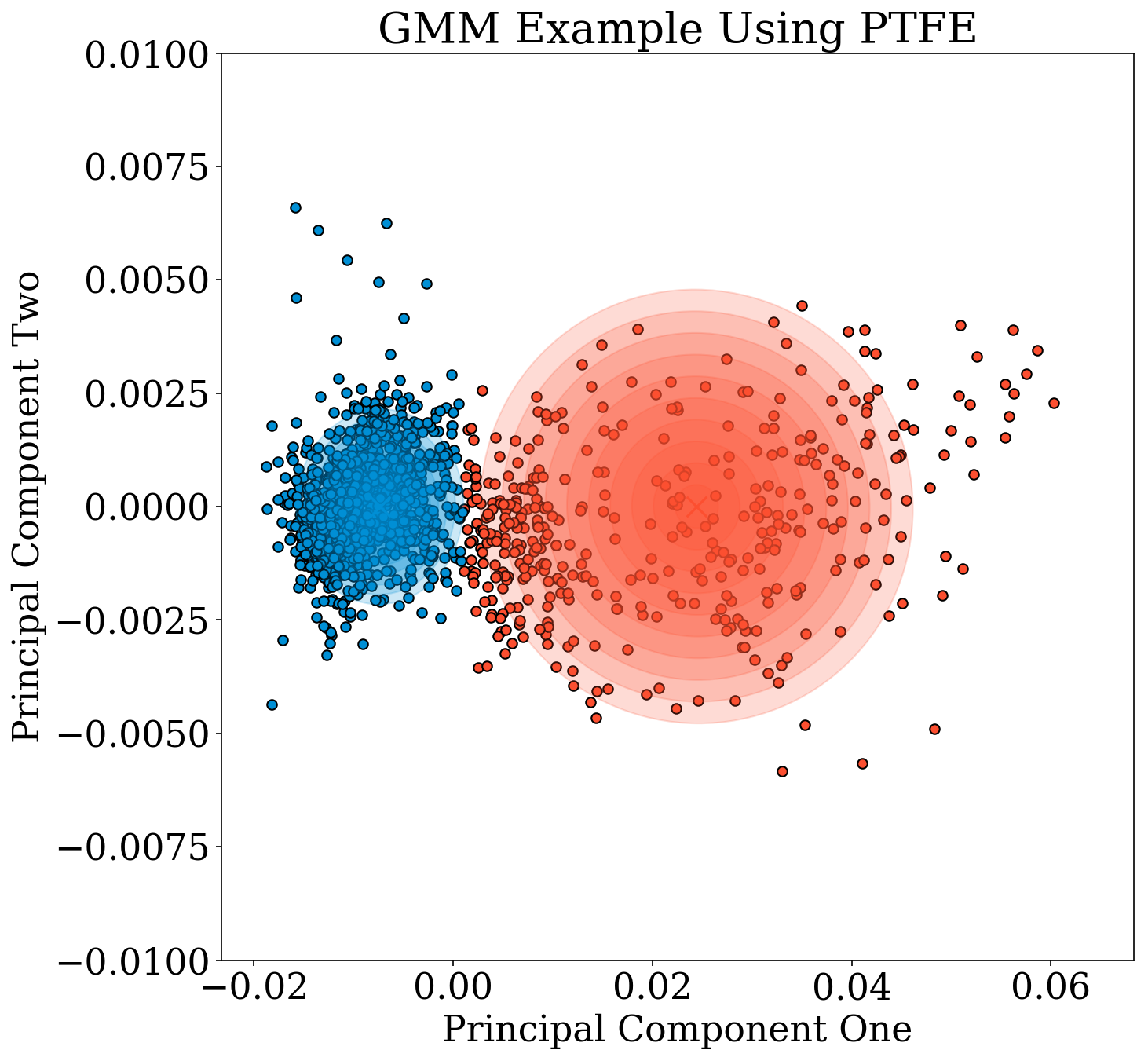}
    \end{subfigure}

    \caption{Comparison of K-means to GMM on the polypropylene image. A contour is displayed representing the  euclidean distance for K-means and the probabililty distribution for the GMM.}
    \label{clustering_methods}
\end{figure}

\subsubsection{Gaussian Mixture Models}

A gaussian mixture model (GMM) is used as a clustering method that implicitly accounts for the Gaussian nature of a dataset. A GMM is expected to perform well due to the Poisson satistics present in x-ray imaging. First we will discuss a general mixture model and then we will see how it is changed by assuming a Gaussian distribution. A more complete overview of GMMs can be found in \cite{DinovExpectationTutorial}.

General mixture models contain $N$ data points, each assumed to be a mixture of $K$ components, with all of these components assumed to be from the same family of distributions, in this case a Gaussian distribution. However these distributions are allowed to have different parameters, such as different means and variances. The model also includes the identity of the mixture component to which each data point belongs. A weighting of each of the $K$ components and sets of parameters for each of the distributions, in this case means and variances for the Gaussians, are necessary. In this work two components were used and an additional constraint was put on the variance of the GMMs to assure that the clusters are spherical in the PCA case.

A bayesian GMM (BGMM) differs from the regular GMM in the fitting of the statistical model to the data. A regular GMM uses the expectation maximization (EM) algorithm to fit the Gaussian to the data. In a BGMM variational inference extends the EM algorithm to maximize the lower bound on model evidence (including priors) instead of the maximum likelihood \cite{Blei2006VariationalMixtures}.

\subsubsection{Other Clustering Methods}

A number of other clustering methods were compared. These additional methods, together with the methods above, are summarized in Table \ref{tab1}.

\begin{table}[htbp]
\caption{Summary of clustering methods}
\begin{tabular}{|l|l|l|l|}
\hline

Method name                                                            & Parameters                                                                                                                                         & Notes on Parameters                                                                                                                & Geometry (metric used)                                                      \\ \hline
K-Means                                                                & number of clusters                                                                                                                                 & \begin{tabular}[c]{@{}l@{}}1 or 2,\\ defined by silhouette\\ score \cite{Rousseeuw1987Silhouettes:Analysis}\end{tabular}                                                    & \begin{tabular}[c]{@{}l@{}}Squared Euclidean\\ Distance\end{tabular}        \\ \hline
Mean-shift \cite{Fukunaga1975TheRecognition}                                                             & bandwidth                                                                                                                                          & \begin{tabular}[c]{@{}l@{}}estimated using\\ scikit-learn's bandwidth\\ estimator\end{tabular}                                     & \begin{tabular}[c]{@{}l@{}}Euclidean \\ Distance\end{tabular}               \\ \hline
Spectral clustering                                                  & \begin{tabular}[c]{@{}l@{}}number of clusters,\\ eigensolver,\\ label assignment\end{tabular}                                                      & \begin{tabular}[c]{@{}l@{}}1 or 2,\\ ARPACK \cite{Lehoucq1998ARPACKMethods},\\ K-means\end{tabular}                                        & \begin{tabular}[c]{@{}l@{}}Nearest-neighbor \\ graph distance\end{tabular}  \\ \hline
\begin{tabular}[c]{@{}l@{}}Ward hierarchical\\ clustering \cite{Ward1963HierarchicalFunction}\end{tabular} & \begin{tabular}[c]{@{}l@{}}number of clusters,\\ connectivity matrix\end{tabular}                                                                  & \begin{tabular}[c]{@{}l@{}}1 or 2,\\ estimated as:\\ $1/2*(C+C')$\\ where C is the\\ K neighbors graph of\\ the data.\end{tabular} & \begin{tabular}[c]{@{}l@{}}Squared\\ euclidean\\ distance\end{tabular}      \\ \hline
HDBSCAN \cite{Campello2015HierarchicalDetection}                                                                & \begin{tabular}[c]{@{}l@{}}minimum samples,\\ minimum cluster size,\\ metric\end{tabular}                                                          & \begin{tabular}[c]{@{}l@{}}10,\\ 10,\\ minkowski\end{tabular}                                                                      & \begin{tabular}[c]{@{}l@{}}Distances between\\ nearest points\end{tabular}  \\ \hline
Gaussian Mixture                                                       & \begin{tabular}[c]{@{}l@{}}number of clusters,\\ covariance type\end{tabular}                                                                      & \begin{tabular}[c]{@{}l@{}}1 or 2,\\ spherical\end{tabular}                                                                        & \begin{tabular}[c]{@{}l@{}}Mahalanobis \\ distances to centers\end{tabular} \\ \hline
Birch \cite{Zhang1996BIRCH}                                               & \begin{tabular}[c]{@{}l@{}}number of clusters,\\ branching factor, \\ threshold\end{tabular}                                                       & \begin{tabular}[c]{@{}l@{}}1 or 2,\\ 19,\\ 0.0001\end{tabular}                                                                     & \begin{tabular}[c]{@{}l@{}}Euclidean \\ distance\end{tabular}               \\ \hline
\begin{tabular}[c]{@{}l@{}}Bayesian \\ Gaussian Mixture\end{tabular}                                                       & \begin{tabular}[c]{@{}l@{}}number of clusters,\\ covariance type,\\ weight concentration\\ prior type,\\ weight concentration\\ prior\end{tabular} & \begin{tabular}[c]{@{}l@{}}1 or 2,\\ spherical,\\ dirichlet process,\\ 1/(number of clusters)\end{tabular}                      & \begin{tabular}[c]{@{}l@{}}Mahalanobis \\ distances to centers\end{tabular} \\ \hline
\end{tabular}

\label{tab1}
\end{table}

A thresholding method, Otsu's thresholding, was also applied to the data. In X-ray image segmentation Otsu's method \cite{Otsu1979AHistograms} is sometimes used for image segmentation \cite{Sund2003AnImaging}. Thus, Otsu's method was also applied as a segmentation method to compare the clustering methods to a more typical approach.

\subsection{Metrics}
\subsubsection{V-measure}

To evaluate the effectiveness of the clustering methods a ground truth was acquired for each image. These ground truths were manually segmented with the help of a reference photograph of the known phantom geometry. For clarity, the labels we give to the ground truth will be referred to as classes while the results of the clustering methods will be clusters.

When comparing the clusters to the ground truth two metrics were considered, homogeneity and completeness: The first metric is homogeneity. Homogeneity is satisfied if a cluster contains data points of a single class. Completeness, satisfied if all data points of a single class are in a single cluster. 

Homogeneity and completeness are formulated mathematically using the method described by Rosenberg and Hirschberg \cite{Rosenberg2007V-Measure:Measure} which can be found in the Appendix. To simplify we build a single metric that uses a combination of both homogeneity $h$ and completeness $c$. This combination is called the V-measure and is the harmonic mean of homogeneity and completeness:

\begin{equation}
V = 2 \cdot \frac{h \cdot c}{h + c}
\end{equation}

This value ranges from 0 to 1, a perfect classification has a score of one while assigning each point to a cluster would result in a score of zero.

\subsubsection{Akaike Information Criterion}

The Akaike information criterion (AIC) \cite{Akaike1998InformationPrinciple} is an estimator of the relative quality of a statistical model for a given data set. Unlike a V-measure, AIC cannot compare different clustering methods, and it was used instead to find parameters for the GMMs. As most clustering models discussed do not include a statistical models the application of AIC is limited to the GMMs.

AIC is a function of the goodness of fit of a statistical model combined with a penalty for overfitting. In this work AIC tested the goodness of fit of the GMM's Gaussian, with a penalty proportional to the number of clusters. Used this way, AIC determines can determine the optimal number of clusters and other GMM parameters such as constraints on the covariance matrix.

A maximum likelihood estimate is used to calculate the AIC. In this case, the maximum likelihood, $\hat L$, was found fitting the GMMs using the EM algorithm and the number of parameters $k$ is the total number of fitted parameters. These values were used to calculate the AIC as:

\begin{equation}
    AIC \, = \, 2k - 2\ln(\hat L)
\end{equation}

For a given model, the lowest AIC denotes a better fit. More specifically an "elbow" or change in curvature of the AIC as a function of the number of components denotes the optimal number of components. For BGMMs, which do not produce a maximum likelihood in training, AIC was calculated using the $\hat L$ from a GMM with the same parameters.

To determine the number of clusters $k$ reliably without human oversight. This was attempted by comparing the AICs and weights for both the $k=1$ and $k=2$ cases. If the AIC did not decrease significantly between one and two clusters and the weights for the two clusters were similar this indicated that there was only one material present.

\begin{wrapfigure}{R}{0.42\textwidth}
  \begin{center}
    \includegraphics[width=0.4\textwidth]{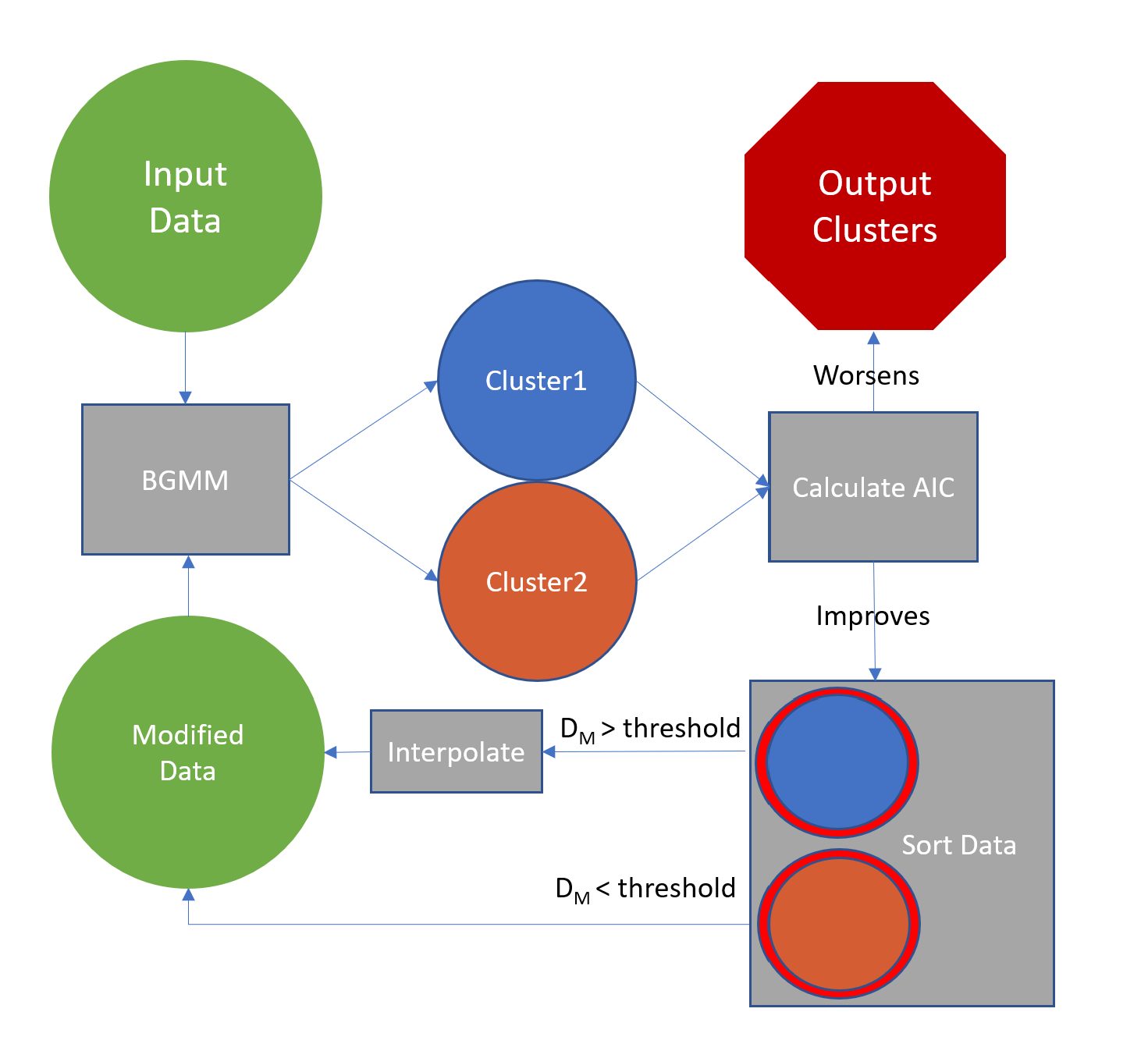}
  \end{center}
        \hspace*{-2in}
    \caption{An overview of the workflow.}
  
  \label{algo_flow}
\end{wrapfigure}

\subsection{Interpolating Bayesian Gaussian Mixture Model}

\begin{figure}[htbp]
    \centering
    \begin{subfigure}[b]{0.48\textwidth}
        \includegraphics[width=\textwidth]{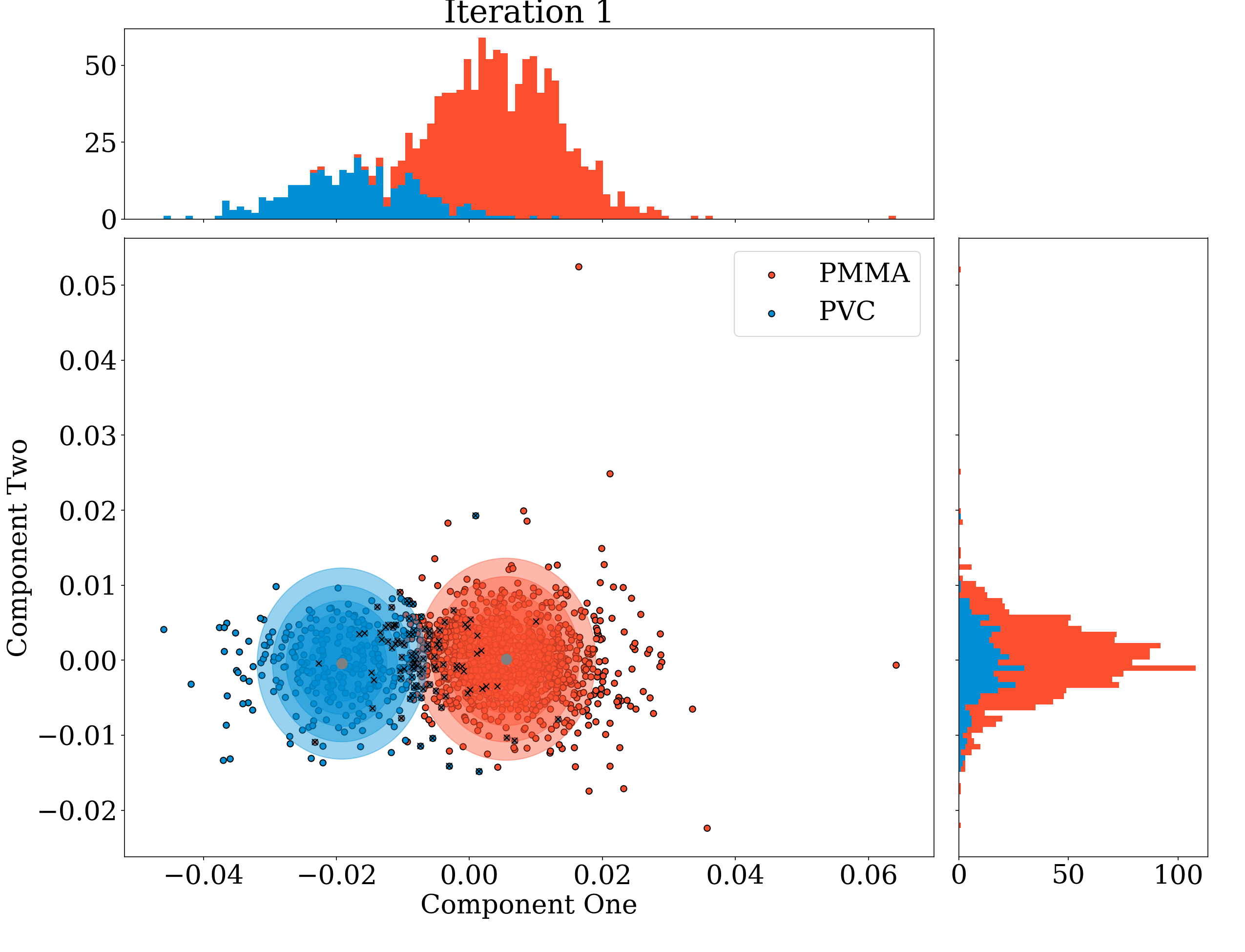}
    \end{subfigure}
    ~ 
    \begin{subfigure}[b]{0.48\textwidth}
        \includegraphics[width=\textwidth]{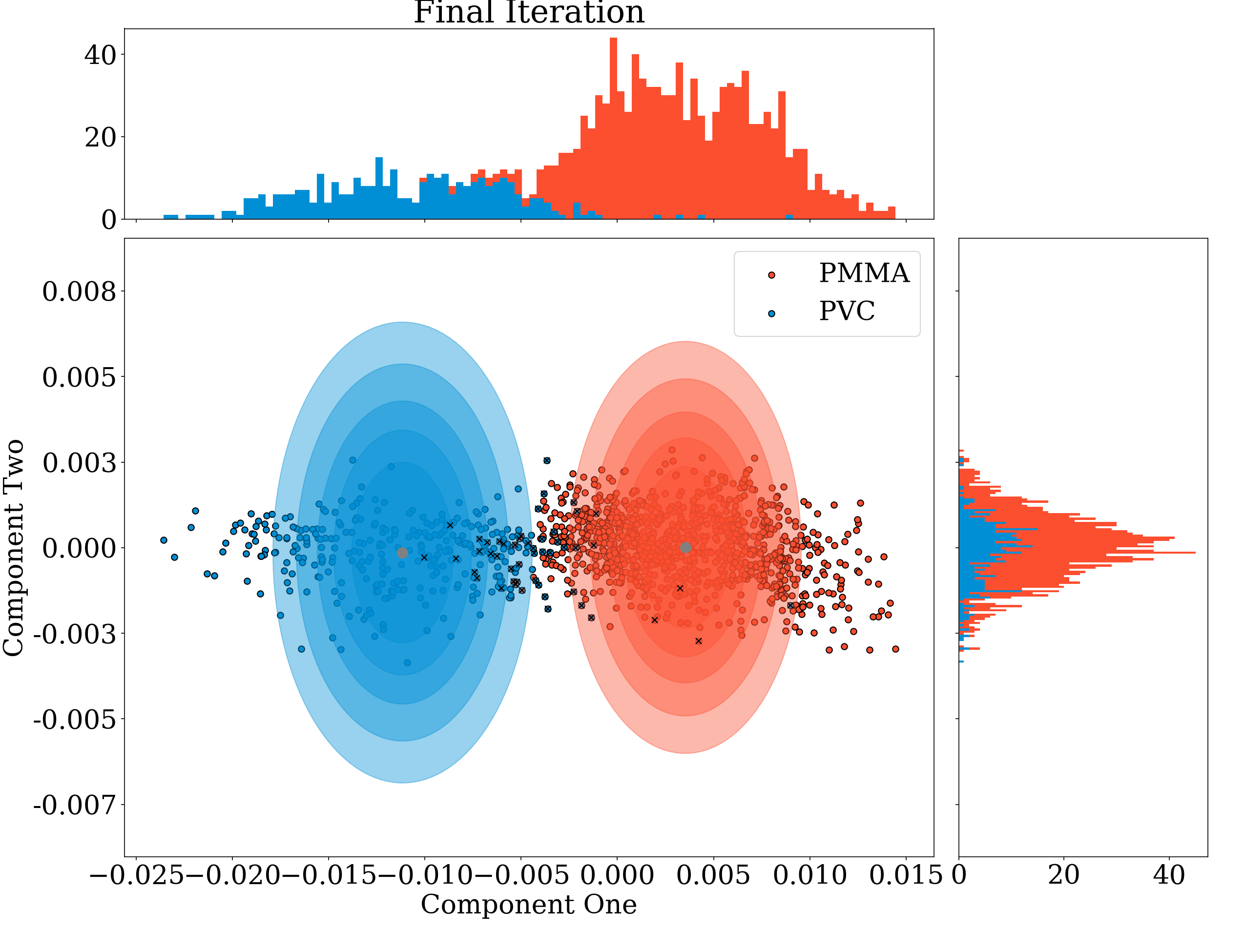}
    \end{subfigure}
    \caption{IBGMM method. The left plot shows the original PVC clustering using the BGMM. The right plot shows the final clustering after convergence. The contours show the contours of the two dimensional Gaussian for each cluster. Black dots are misclassified pixels.}
    \label{iterative_method}
\end{figure}

Additional to the previous methods, a new clustering method was also proposed. The workflow of this method is seen in Figure \ref{algo_flow}.  A demonstration of this method can be seen in Figure \ref{iterative_method}. A BGMM was trained as in the GMM section above. Within each cluster the Mahalanobis distance \cite{Mahalanobis1936OnStatistics} from all the points to the center was calculated. Points at a distance greater than 90\% of the maximum distance were then identified as potential outliers. Each potential outlier was then interpolated spatially using biharmonic interpolation \cite{Damelin2017OnAspects}. The BGMM was then retrained using starting point defined by the end state of the previous iteration. The process was then repeated until convergence which was indicated by an increasing AIC, typically 5-10 iterations.

\subsection{Post-processing}

\begin{wrapfigure}[15]{R}{0.48\textwidth}
  
  \begin{center}
    \includegraphics[width=0.48\textwidth]{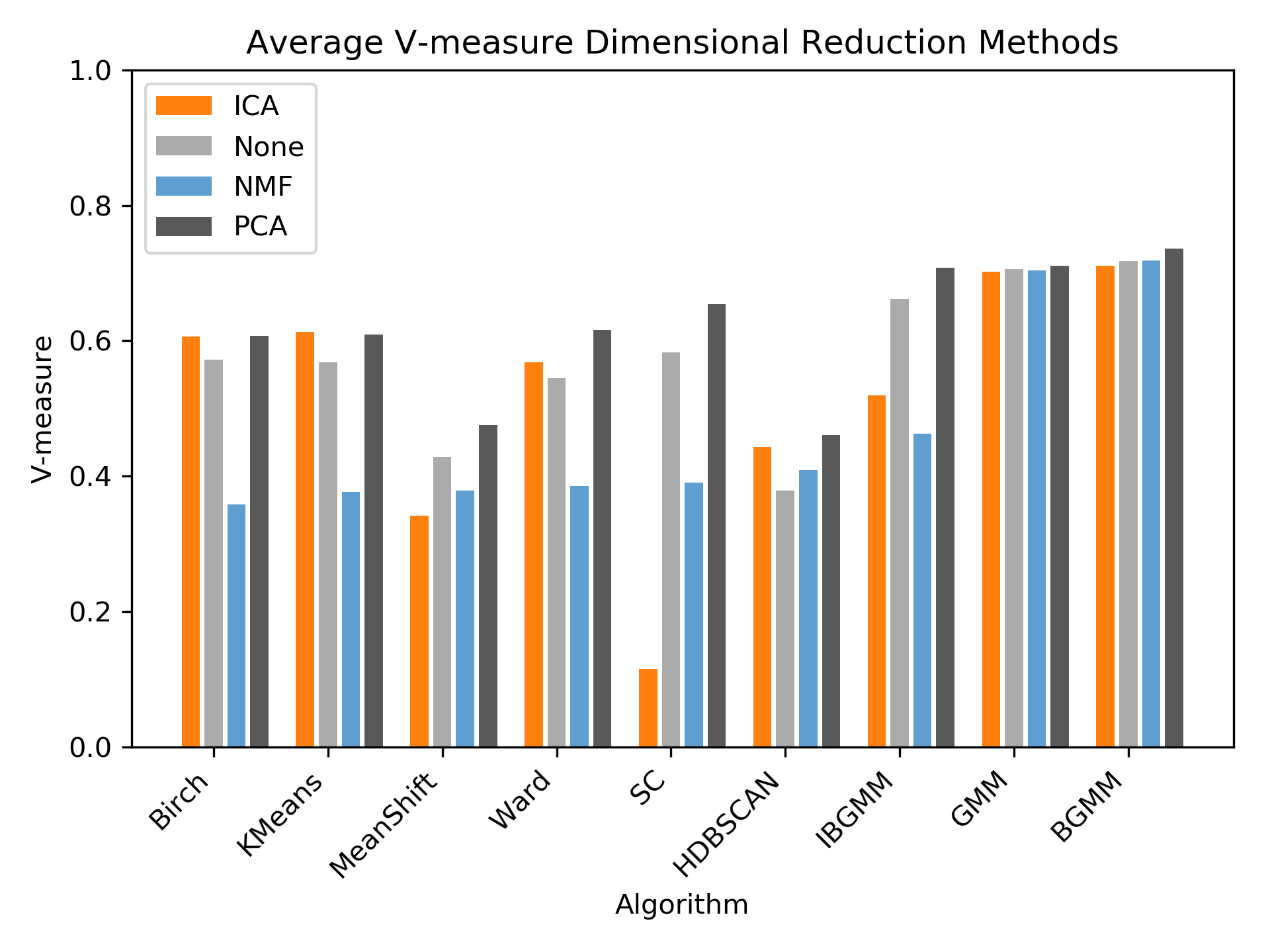}
  \end{center}
  
  \caption{Dimensional reduction results}
  
  \label{results:dr}
\end{wrapfigure}

The aim of this work was to entirely automate image segmentation. It was therefore necessary to introduce an identification layer. This layer identified if there were two materials present in the segmented image after clustering was applied using the following framework:

First a uniform filter was applied to the image so that the binary data becomes continuous. This acts to reduce the score of isolated pixels that have been identified. Secondly, we apply a threshold to the image of 0.2. Next, we fill all of the binary holes in the image to make candidates for segmentation more uniform. A bounding box is then created around all of the nonzero elements of the image and overlaid on the image as the final output.


\section{Results and Discussion}

\subsection{Dimensional Reduction}

Figure \ref{results:dr} shows the V-measures for different dimensional reduction methods averaged over all materials for each clustering method. For comparison the results without dimensional reduction are also displayed.

Overall PCA had the best results. V-measures were 0.71, 0.72, 0.72, and 0.74 for ICA, no dimensional reduction, NMF, and PCA respectively. PCA was on average 2.7\% better than ICA and 2.6\% better than NMF in the best case. All other clustering methods performed best with PCA other than K-means which performed better with ICA.


\begin{figure}[t!]
    \centering
    \begin{subfigure}[b]{0.32\textwidth}
        \includegraphics[width=\textwidth]{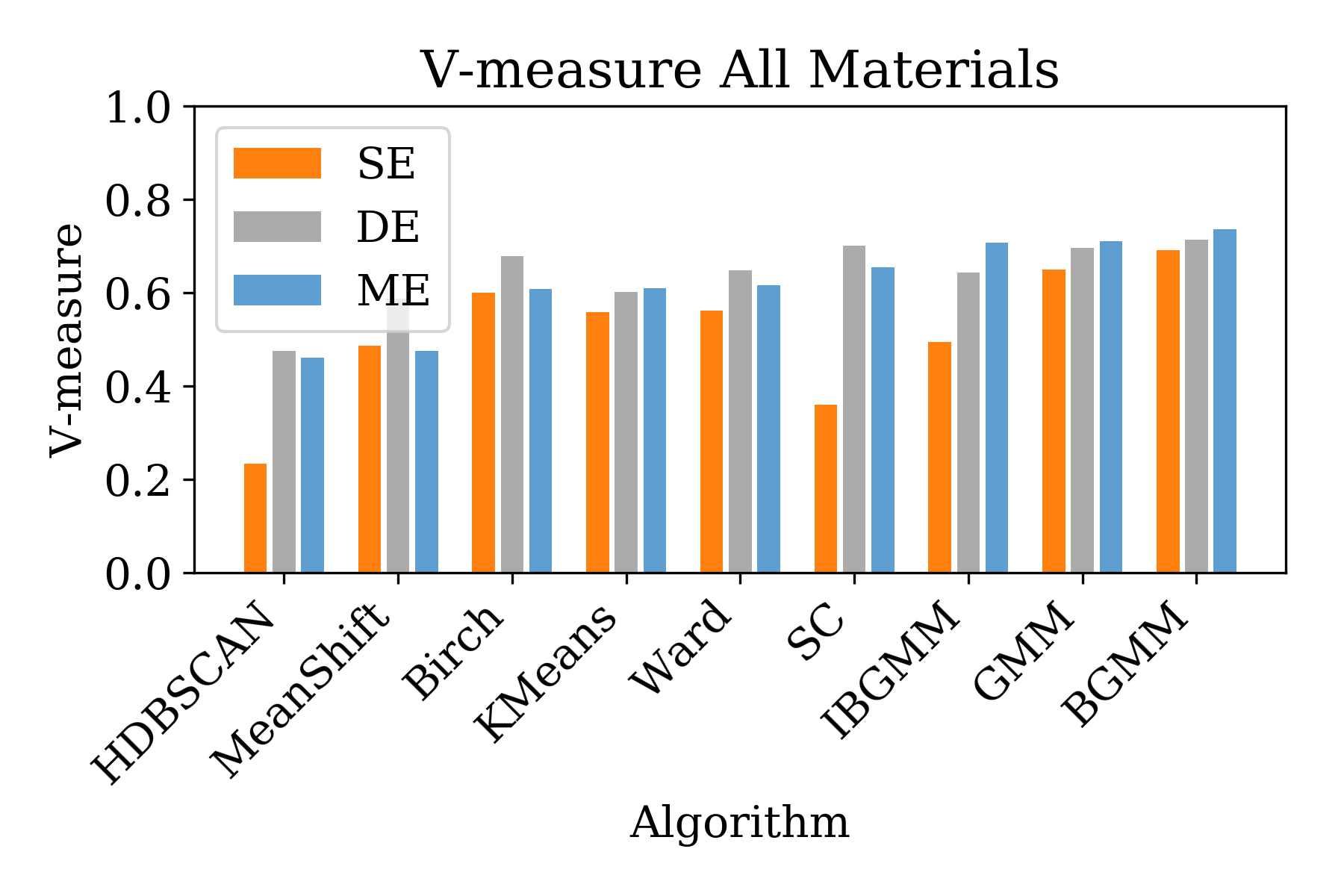}
    \end{subfigure}
    ~ 
    \begin{subfigure}[b]{0.32\textwidth}
        \includegraphics[width=\textwidth]{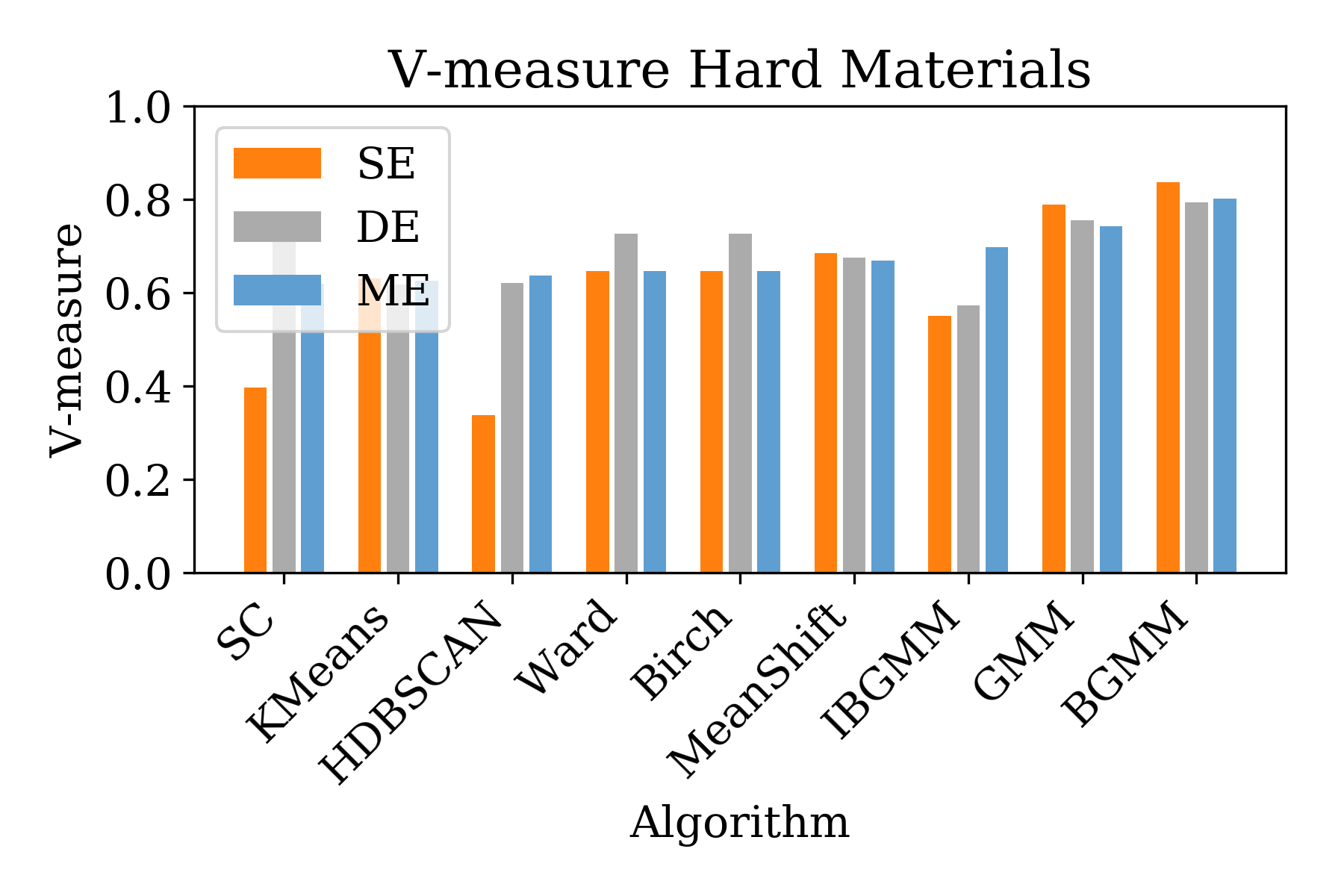}
    \end{subfigure}
    \begin{subfigure}[b]{0.32\textwidth}
        \includegraphics[width=\textwidth]{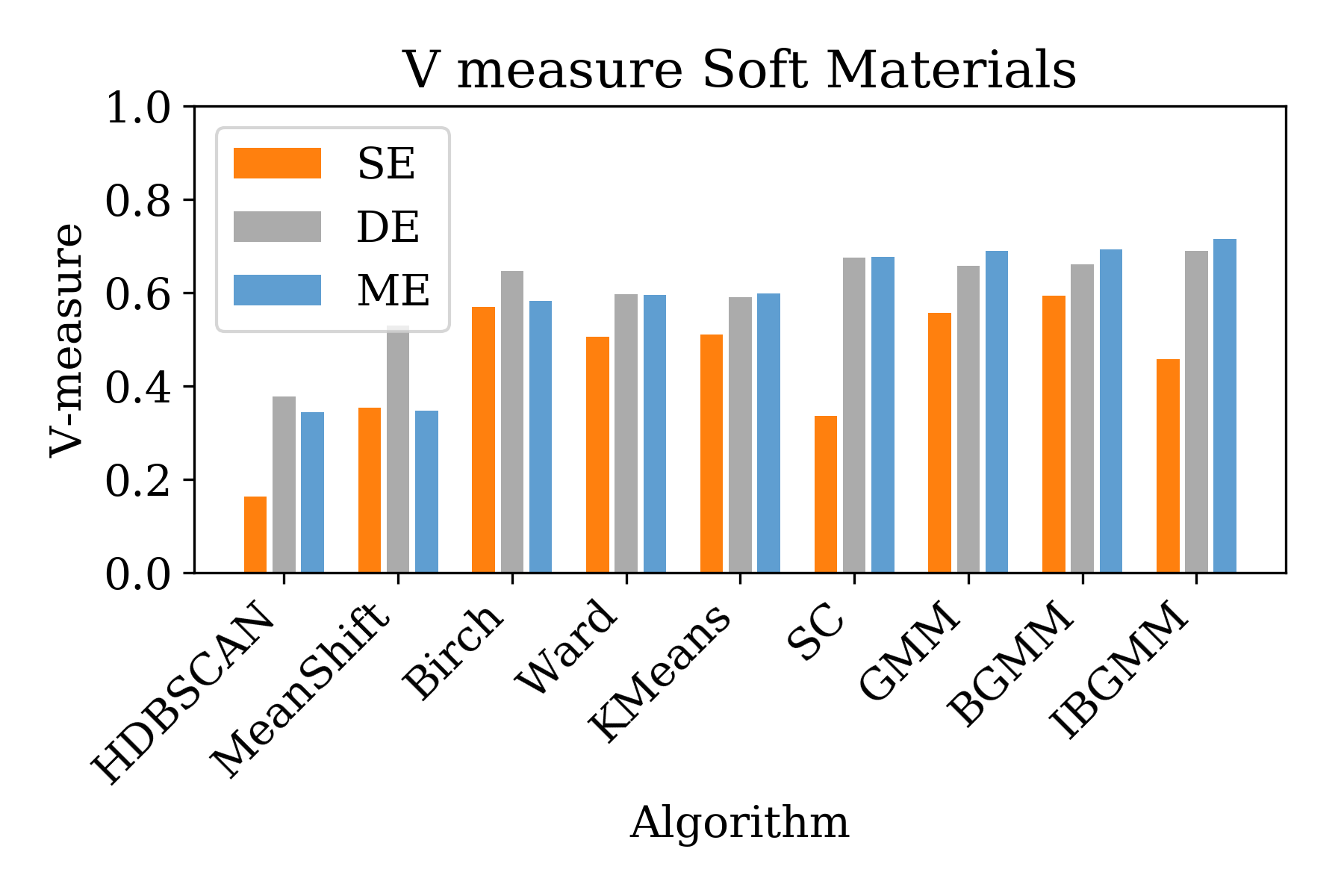}
    \end{subfigure}
    \caption{Comparison to Dual and Single Energy using V-measure with results averaged over all of the tissues, soft tissues and hard tissues respectively}
    \label{results:hard_soft}
\end{figure}

\subsubsection{Single, Dual, and Spectral Comparison}

Overall ME imaging was seen to be marginally better that DE and SE imaging when the results were averaged over all investigated materials (Figure \ref{results:hard_soft}. The V-measure for the BGMM method which was best in all modalities was 0.74, 0.71 and 0.69 for ME, DE and SE x-ray imaging respectively. V-measure of ME imaging was on average 4.2\% better than DE imaging and 7.2\% better than SE imaging.

Figure \ref{results:hard_soft} b-c) show the results separated in terms of the material density. Hard materials such as glass and steel had the best results for SE clustering with BGMM. The best V-measures for the hard tissues were 0.80, 0.79, and 0.84 for ME, DE and SE respectively. While for soft materials (PVC, PTFE, PP) the IBGMM method provided the best results when clustering with ME and DE imaging with V-measures of 0.71 and 0.69 respectively, SE with BGMM performed best with a V-measure of 0.59. This was a gain of 16.9\% between SE and DE and a gain of 3.5\% between DE and ME.

\subsubsection{Number of Clusters}

\begin{figure}[htbp]
    \centering
    \begin{subfigure}[b]{0.48\textwidth}
        \includegraphics[width=\textwidth]{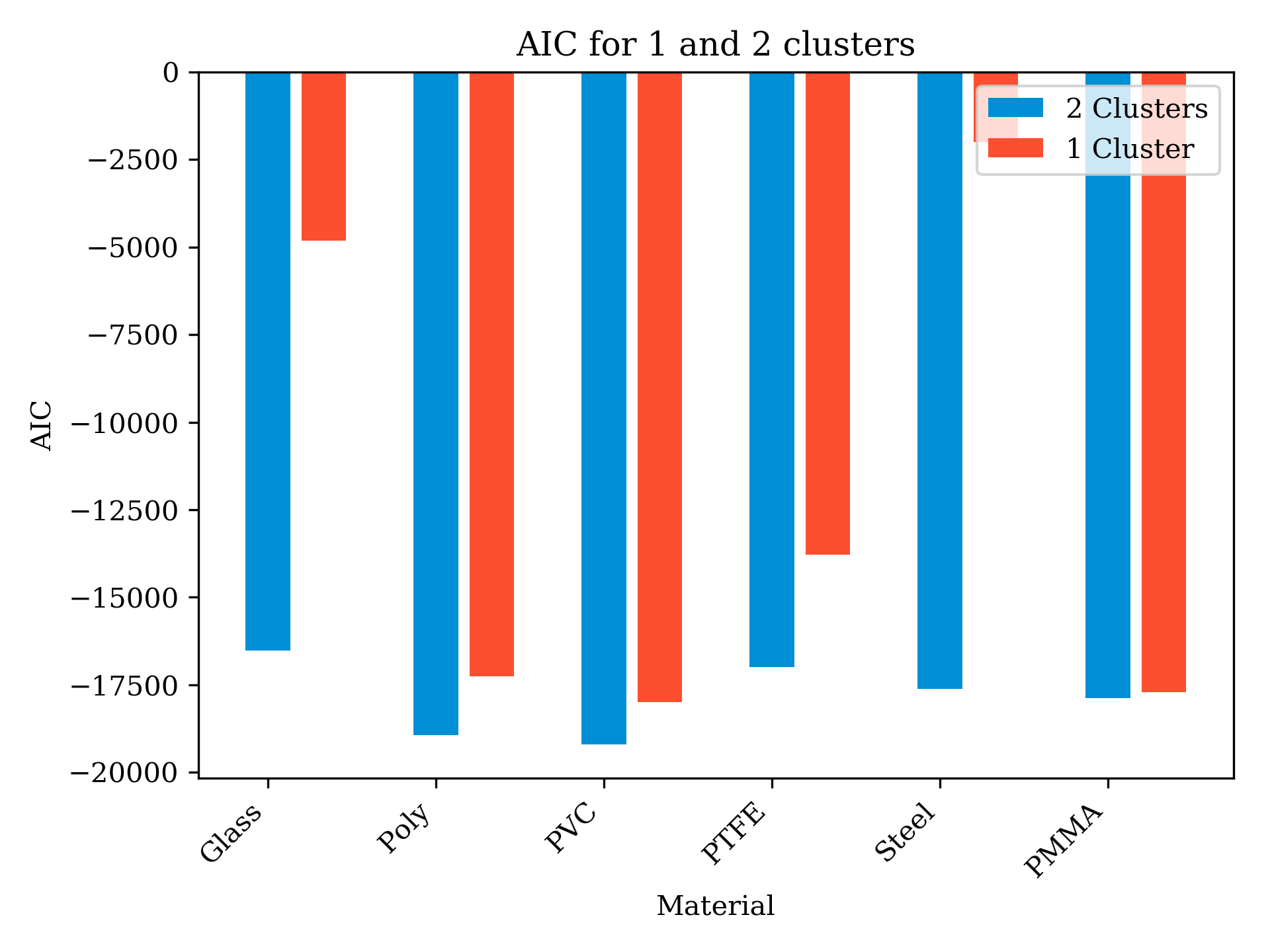}
    \end{subfigure}
    \begin{subfigure}[b]{0.48\textwidth}
        \includegraphics[width=\textwidth]{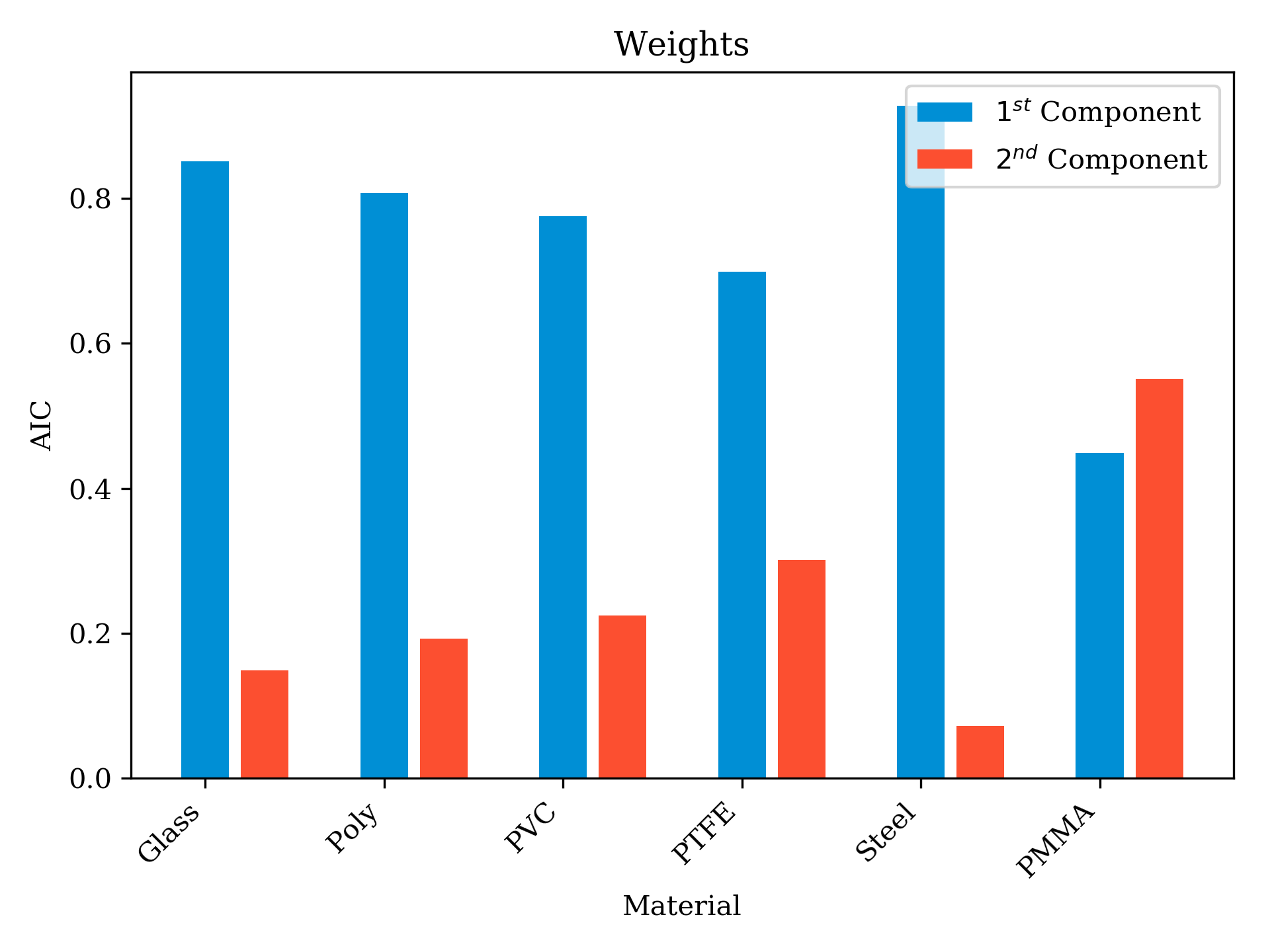}
    \end{subfigure}
    \caption{The comparison in AIC (a) and weights (b) for 1 and 2 clusters for the investigated materials as well as a PMMA only scan.}
    \label{n_bins}
\end{figure}

The results for the multi-material phantom as well as for a blank scan of PCA are shown below. The AIC for all of materials was seen to decrease by on average 6683 and 1208 in the worst case. For the blank scan the AIC was seen to decrease by 160. The weights for the different materials were seen to be in the most equal case \%40 different for PTFE and was only \%10 different in the one material case as seen in figure \ref{n_bins}.

\subsubsection{Coparison to Threshold Method}

\begin{wrapfigure}[44]{r}{0.33\textwidth}
  
  \begin{center}
    \includegraphics[width=0.33\textwidth]{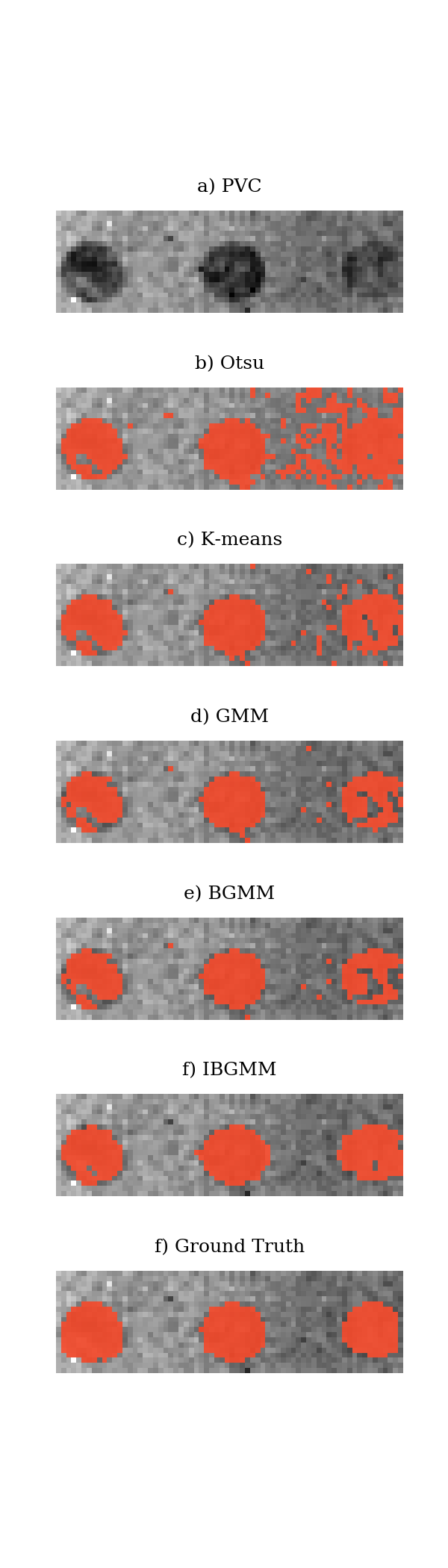}
  \end{center}
  
  \caption{Overlay of the different segmentation methods on the PVC image.}
  
  \label{results:otsu}
\end{wrapfigure}

As can be seen in Figure \ref{results:otsu} when the background had similar composition to the material in question there was erratic thresholding of the background. In this case the IBGMM had a V-measure of 0.74 while the thresholding method has a V-measure of 0.41. One can see in Figure \ref{results:otsu} that thresholding cannot resolve the thinnest (2mm) insert in the image and the IBGMM is the only method that is able to recover the circular shape of the insert while suppressing noise in the data.

\begin{table}[b!]
\caption{Use Cases}
\label{table:results}
\begin{tabular}{llll}
   & Recommended Use Case       & Recommended method & V-measure (hard | soft) \\ \hline
ME & Soft material segmentation & IBGMM with PCA     & 0.71 | 0.80             \\
DE & -                          & IBGMM with PCA     & 0.69 | 0.79             \\
SE & Hard material segmentation & BGMM               & 0.59 | 0.84            
\end{tabular}
\end{table}


\section{Discussion}

ME imaging was seen to have a 20.3\% higher V-measure than SE and 3.5\% higher V-measure than DE imaging when segmenting soft materials. Conversely, SE imaging was more effective by both DE and ME imaging for the hard material segmentation with a margin of 6.3\% and 5.0\% respectively. The best methods and results are summarized in Table \ref{table:results}. These results are consistent with results from ME CT, Lalonde \cite{Lalonde2016ACT} found that three energies reduced errors in elemental composition calculations, but only marginally compared to the gains between SE and DE imaging. Likewise in this application, SE imaging is seen to be much less effective than DE imaging for soft material segmentation while DE and ME have similar results with ME having a 3.5\% higher V-measure. This raises the question as to if ME or DE imaging are ideal for this application? Moreover, are the considerable downsides of ME and DE imaging worth the potential benefits?

Although prices continue to decrease, ME detectors are generally more expensive compared to photon-integrating detectors. To validate the benefit for the additional cost, one would need an application that has a high throughput, is facilitated by a small detector, and needs an accurate segmentation of soft tissues. A potential candidate that fits these requirements would be mammography, since mammography machines have high throughput, image a relatively small area and aim to segment soft tissues.

This work discusses unsupervised methods for computer vision, recently supervised methods have produced superior results to unsupervised methods in many areas such as brain segmentation \cite{Bakas2018IdentifyingChallenge}, and general tumor segmentation \cite{Simpson2019AAlgorithms}. So why implement these methods rather than a deep learning method?

Deep learning has taken medical imaging by storm. In recent years much success has been had implementing deep learning image segmentation \cite{IsenseeNnU-Net:Segmentation,Ronneberger2015U-Net:Segmentation}. However, deep learning implementations include considerable downfalls \cite{Marcus2018DeepAppraisal}. These methods are only as good as their training data, and deep learning models have no guarantee to converging to actual system modelled. A gradient descent algorithm may converge to a local minimum or overfit the data. Overfiffing introduces a chance of the model behaving unpredictably when new input data is used. This puts a large pressure on having a large and varied training dataset which is the best way to avoid overfitting.

Imaging modalities are constantly improving. The equipment of five or ten years ago often has different capabilities than the state of the art today. Thus, having a machine learning system that is dependant on clinical data for training goes contrary to the development cycle of imaging modalities. Large clinical datasets will not be acquired until the modality has been adopted by a large amount of health care providers. This begs the question; if we are to implement new imaging modalities to leverage deep learning how do we validate their utility without large clinical datasets?

Thus this work offers an alternate approach. Although deep learning segmentation \cite{Ronneberger2015U-Net:Segmentation} would likely result in superior segmentation. A clinical dataset for ME imaging does not yet exist, but this could change, if ME imaging is introduced into clinical practice in coming years. Thus these methods could exist in the interim, between when ME imaging is implemented and before large clinical datasets become available. Further, having an automated segmentation method would be of particular importance in clinical ME imaging since manual segmentation is not as effective: Clinicians are disadvantaged compared to a computer when visualizing five dimensional images as compared to one dimensional images.

Some of the clustering methods this application is not ideal, however, these methods have been left in the analysis for completeness. Particularly, the mean-shift and HDBSCAN methods, both performing poorly, do not include the parameter $k$ and thus define what they consider clusters in the data. Since the phantom is made of three inserts of different thicknesses there is reason for a method to fit more than two clusters to the data. Thus, it is not unexpected that these methods preform poorly in this task as we have not given them as much information as the other methods. 


It was seen that PCA is the optimal dimensional reduction method for ME imaging in this task. Although ICA shows promise in hyperspectral imaging, it did not show improvement over the unreduced data in this implementation. In the hyperspectral implementation domain knowledge was used to calculate the exact spectral response of the materials to be segmented. Currently we were not able to model the spectral response of the materials and thus could not apply this domain knowledge. With an analytical model of the CZT detector response and a ray tracer one could calculate the mean attenuation for each energy bin. These theoretical mean values could then be compared to the mean value of a GMM's cluster to identify a material.

It was observed that using the AIC and weights from the GMMs the parameter $k$ could be tuned for an algorithm. A scan of pure PMMA showed a difference of 160 between the AIC for $k=1$ and $k=2$ while all scans that included more than one material produced a difference in the AIC of at least 6683. Further, in the pure PMMA scan, weights between the two materials were seen to be similar with only a 10\% difference. These two metrics, in combination, can be used to avoid false negatives if image contains only one material.

For SE imaging the BGMM still showed the best results in this application. This approach could be used as an effective automatic segmentation technique superior to windowing for soft tissue segmentation. For ME and DE the IBGMM approach could yield superior segmentations and could see applications in breast lesion detection. IBGMM could serve as an automated second look working in tandem with a clinical professional. Further work could also investigate a semi-automated workflow, working with a trained person to cluster using user selected regions to be used as the mean of the GMM, letting the user define the number of clusters and select materials they would like the clusters to include.

\section{Conclusion}

In summary, segmentation of ME, DE, and SE data has been presented as well as an improved method for image segmentation of soft materials. It was seen that ME imaging had the highest V-measure on the soft materials using PCA and a novel IBGMM method with a V-measure of 0.71. This method was 3.5\% better than DE and 20.3\% better than SE. Conversely, SE imaging is most capable of hard tissue segmentation using a BGMM having the highest V-measures of 0.84. This method was 5.0\% better than ME and 6.3\% better than DE on the same task.


\section*{Acknowledgments}

The authors would like to thank Hugh Patterson for manufacturing the imaging phantom. NSERC, Redlen Technologies and the Canada Research Chairs Program are acknowledge for funding support of this project.

\appendix{}
\section{Mathematical Derivations}

\subsection{V-measure}

To give a metric to describe the disorder in our clustering result we use the Shannon entropy of the classes given the clustering assignment $H(C|K)$ and the entropy of the classes  $H(C)$.

\begin{equation}
H(C|K) = - \sum_{c=1}^{|C|} \sum_{k=1}^{|K|} \frac{n_{c,k}}{n}
\cdot \log\left(\frac{n_{c,k}}{n_k}\right)
\end{equation}

\begin{equation}
H(C) = - \sum_{c=1}^{|C|} \frac{n_c}{n} \cdot \log\left(\frac{n_c}{n}\right)
\end{equation}

with $n$ the total number of data points, $n_c$ and $n_k$ the number of data points belonging to class $c$ and cluster $k$ respectively, and $n_{c,k}$ as the number of data points from class $c$ assigned to cluster $k$.

Rosenberg and Hirschberg then define the homogeneity $h$ and completeness $c$ as:

    



In this study, methods were evaluated according to a combination of both homogeneity $h$ and completeness $c$. This combination is called the V-measure and is the harmonic mean of homogeneity and completeness.

\begin{equation}
V = 2 \cdot \frac{h \cdot c}{h + c}
\end{equation}

\subsection{ICA}

The components $x_i$ of the images with 5 bins $\boldsymbol{x}=(x_1,\ldots,x_5)^T$ are seen to be a sum of the independent components $s_k$, $k=1,\ldots,5$:

$x_i = a_{i,1} s_1 + \cdots + a_{i,k} s_k + \cdots + a_{i,5} s_5$

where $a_{i,k}$ are the mixing weights.

Or in matrix form as $\boldsymbol{x}=\sum_{k=1}^{5} s_k \boldsymbol{a}_k$, where our image vectors $\boldsymbol{x}$ are represented by the basis vectors $\boldsymbol{a}_k=(\boldsymbol{a}_{1,k},\ldots,\boldsymbol{a}_{m,k})^T$. The basis vectors $\boldsymbol{a}_k$ form the columns of the mixing matrix $\boldsymbol{A}=(\boldsymbol{a}_1,\ldots,\boldsymbol{a}_5)$.

Putting all this together we have the matrix equation $\boldsymbol{x}=\boldsymbol{A} \boldsymbol{s}$, where $\boldsymbol{s}=(s_1,\ldots,s_5)^T$.

Given our images $\boldsymbol{x}_1,\ldots,\boldsymbol{x}_N$ of the random vector $\boldsymbol{x}$, the task is to estimate both the mixing matrix $\boldsymbol{A}$ and the sources $\boldsymbol{s}$. This is done by adaptively calculating the $\boldsymbol{w}$ vectors and setting up a cost function which maximizes the non-gaussianity of the calculated $s_k = \boldsymbol{w}^T \boldsymbol{x}$. In this work 
the maximum likelihood estimate (MLE) algorithm was used for finding the unmixing matrix $W$.

\bibliographystyle{JHEP}
\bibliography{oconnell_2019}

\vfill


\end{document}